\documentclass[aps,superscriptaddress,twocolumn,twoside,floatfix,pra,a4paper]{revtex4-1} %nofootinbib
\usepackage{times}
\usepackage{epsfig}
\usepackage{amsfonts}
\usepackage{amsmath}
\usepackage{amssymb}
\usepackage{color}
\usepackage{multirow}

\usepackage{latexsym}
\usepackage{amsfonts}
\usepackage{mathrsfs}
\usepackage{natbib}
\usepackage{verbatim}
\usepackage{gensymb}

\usepackage[colorlinks=true,linkcolor=blue,citecolor=magenta,urlcolor=blue]{hyperref}

\DeclareMathOperator{\Tr}{Tr}

\newcommand{\ket}[1]{|#1\rangle}
\newcommand{\bra}[1]{\langle#1|}
\newcommand{\braket}[2]{\langle#1|#2\rangle}
\newcommand{\bracket}[3]{\langle#1|#2|#3\rangle}
\newcommand{\ketbra}[2]{|#1\rangle\langle#2|}

\newcommand{\ba}{\begin{eqnarray}}
\newcommand{\be}{\begin{equation}}
\newcommand{\ee}{\end{equation}}

\newcommand{\ea}{\end{eqnarray}}
\newcommand{\ban}{\begin{eqnarray*}}
	\newcommand{\ean}{\end{eqnarray*}}

\newcommand{\figref}[1]{Fig.~\ref{#1}}

\newcommand{\appref}[1]{App.~\ref{#1}}

\begin{document}
	
\title{Autonomous multipartite entanglement engines}

\author{Armin Tavakoli}
\affiliation{Department of Applied Physics, University of Geneva, 1211 Geneva, Switzerland}

\author{G\'eraldine Haack}
\affiliation{Department of Applied Physics, University of Geneva, 1211 Geneva, Switzerland}

\author{Nicolas Brunner}
\affiliation{Department of Applied Physics, University of Geneva, 1211 Geneva, Switzerland}

\author{Jonatan Bohr Brask}
\affiliation{Department of Applied Physics, University of Geneva, 1211 Geneva, Switzerland}
\affiliation{Department of Physics, Technical University of Denmark, Fysikvej, 2800 Kongens Lyngby, Denmark}

\begin{abstract}
The generation of genuine multipartite entangled states is challenging in practice. Here we explore a new route to this task, via autonomous entanglement engines which use only incoherent coupling to thermal baths and time-independent interactions. We present a general machine architecture, which allows for the generation of a broad range of multipartite entangled states in a heralded manner. Specifically, given a target multiple-qubit state, we give a sufficient condition ensuring that it can be generated by our machine. We discuss the cases of Greenberger-Horne-Zeilinger, Dicke and cluster states in detail. These results demonstrate the potential of purely thermal resources for creating multipartite entangled states useful for quantum information processing.
\end{abstract}

\date{\today}

\maketitle

%%%%%%%%%%%%%%%%%%%%%%%%%%%%%%%%%%%%%%%%%%%%%%%%%%%%%%%%%%%%%%%%%%%
% Main text
%%%%%%%%%%%%%%%%%%%%%%%%%%%%%%%%%%%%%%%%%%%%%%%%%%%%%%%%%%%%%%%%%%%

\textit{Introduction.---}Quantum thermal machines combine quantum systems with thermal reservoirs at different temperatures and exploit the resulting heat flows to perform useful tasks. These can be work extraction or cooling, in analogy with classical heat engines and refrigerators, but may also be of a genuinely quantum nature. In particular, it is possible to devise entanglement engines -- thermal machines generating entangled quantum states. Entanglement is a key resource for quantum information processing but is generally very fragile and easily destroyed by environmental noise. It is nevertheless possible to exploit dissipation to create and stabilise entanglement \cite{Plenio1999,Plenio2002,Schneider2002,Kim2002,Jakobczyk2002,Braun2002,Benatti2003,Hartmann2006,Quiroga2007,Burgarth2007,Kraus2008,Diehl2008,Verstraete2009}. This was studied in a variety of settings and physical systems \cite{Cai2010,Kastoryano2011,Znidaric2012,Bellomo2013,Reiter2013,Schuetz2013,Walter2013,Ticozzi2014,Boyanovsky2017,Hewgill2018,Lee2019} and dissipative entanglement generation using continuous driving was experimentally demonstrated, mainly for bipartite states \cite{Krauter2011,Barreiro2011,Shankar2013,Lin2013}. 

Autonomous entanglement engines represent a particularly simple case. Here, entanglement can be generated dissipatively with minimal resources, using only time-independent interactions and contact to thermal reservoirs at different temperatures. No driving, coherent control, or work input is required. For the bipartite case, a two-qubit entangled state can be generated in a steady-state, out-of-thermal-equilibrium regime \cite{Brask2015}. Although the entanglement produced by such machines is typically weak, it can be boosted via entanglement distillation \cite{Bennett1996}, or by coupling to negative-temperature \cite{Tacchino2018} or joint baths \cite{Man2019}. In fact, applying a local filtering operation to the steady state of a bipartite entanglement engine can herald maximal entanglement between two systems of arbitrary dimension \cite{Tavakoli2018}. 

These first results show that using dissipative, out-of-equilibrium thermal resources offers an interesting perspective on entanglement generation. A natural question is whether this setting could also be used to generate more complex forms of entanglement, in particular entanglement between a large number of subsystems. It is of fundamental interest to understand the possiblities and limits of thermal entanglement generation. In addition, such multipartite entangled states represent key resources, e.g. for measurement-based quantum computation, quantum communications, and quantum-enhanced sensing and metrology. The creation and manipulation of complex entangled states is therefore of strong interest for many experimental platforms, although typically very challenging in practice. 

Here, we propose autonomous entanglement engines as a new route to the generation of multipartite entanglement and explore their potential. A first question is, which types of multipartite entangled states can be created. We present a sufficient condition for a given target $N$-qubit state to be obtainable. Specifically, for any target state satisfying our criterion, we construct an autonomous entanglement engine that will generate this state. The engine consists of $N$ interacting qutrits (three-level systems), each qutrit being locally connected to a thermal bath. From the resulting steady state, a local filtering operation then leads to the desired target state. In particular, our scheme can generate important classes of genuine multipartite entangled states, including Greenberger-Horne-Zeilinger (GHZ), Dicke and cluster states, which we discuss in detail. We show that these states can be generated with high fidelities and good heralding probabilities.

\begin{figure}[t]
	\begin{center}
		\includegraphics[width=\linewidth]{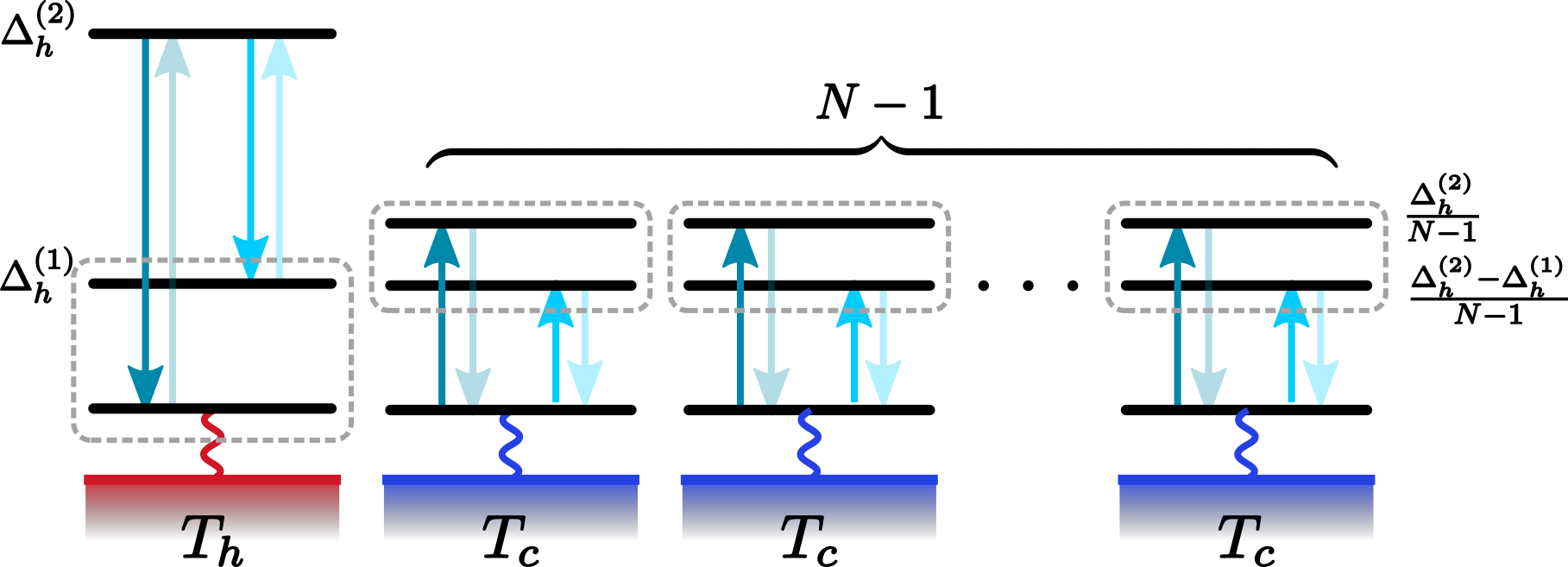}
		\caption{Autonomous thermal machine for the generation of $N$-qubit GHZ states. One qutrit is coupled to a hot thermal bath, while $N-1$ qutrits are coupled to cold thermal baths at equal temperatures. The energy level structure (not drawn to scale) is such that transitions in the hot qutrit are resonant with collective transitions of the cold qutrits, as indicated by arrows. All the cold systems have the same structure, i.e.~$\Delta_k^{(1)} = \Delta_c^{(1)}$ and $\Delta_k^{(2)} = \Delta_c^{(2)}$  for $k=2,\ldots,N$, and $\Delta_c^{(1)} = (\Delta_h^{(2)}-\Delta_h^{(1)})/(N-1)$ and $\Delta_c^{(2)} = \Delta_h^{(2)}/(N-1)$. Local filters, when successful, project the qutrits onto the qubit subspaces enclosed in dashed, gray boxes. }
		\label{fig.ghzmachine}
	\end{center}
\end{figure}

\textit{Entanglement engine.---}We begin by describing the entanglement engine. The structure of the machine is determined by the choice of subspace, energy spectrum, and bath temperature for each qutrit, as well as the form of the interaction, all of which generally depend on the $N$-qubit target state $\ket{\psi}$. This state is obtained in a heralded manner from the steady state of the machine by projection of each qutrit to a qubit subspace. \figref{fig.ghzmachine} shows an example targeting a GHZ state.

The machine evolution consists of a Hamiltonian contribution and a dissipative contribution due to the heat baths. The evolution is autonomous in the sense that both the Hamiltonians and the bath couplings are time independent, and the machine thus requires no work input to run. Denoting the energy basis states of qutrit $k$ by $\{\ket{0}_k$, $\ket{1}_k$, $\ket{2}_k\}$ and taking the corresponding energies to be $\{0, \Delta_k^{(1)}, \Delta_k^{(2)}\}$, the free Hamiltonian of each qutrit is $H_k = \Delta_k^{(1)} \ket{1}_k\bra{1} + \Delta_k^{(2)} \ket{2}_k\bra{2}$. The free Hamiltonian of the machine is
\begin{equation}
\label{Hfree}
H_{\text{free}} = \sum_{k=1}^N H_k = \sum_{k=1}^N\left(\sum_{l=1}^{2} 	\Delta_k^{(l)}\ket{l}_k\bra{l}\right) .
\end{equation}
In addition, the qutrits interact via a time-independent Hamiltonian $H_{\text{int}}$, specified below.

We model the machine evolution including the heat-bath induced dissipation with a master equation of the form 
\begin{equation}
\label{mastereq}
\frac{d\rho}{dt} = -i [H_{\text{free}}+H_{\text{int}} , \rho] + \mathcal{L}(\rho) .
\end{equation}
For simplicity, we adopt a local reset model in which the dissipator $\mathcal{L}$ corresponds to spontaneous, probabilistic, independent resets of each qutrit to a thermal state at the corresponding temperature \cite{Hartmann2006,Linden2010}. That is,
\begin{equation}
\label{Lreset}
\mathcal{L}(\rho) = \mathcal{L}_k(\rho) = \sum_{k=1}^{N} \gamma_k (\tau_k \otimes_{k} \Tr_{k} (\rho) - \rho) .
\end{equation}
where $\gamma_k$ is the reset rate for qutrit $k$, $\tau_k = \exp(-H_k/T_k) / \Tr[\exp(-H_k/T_k)] $ is a thermal state of qutrit $k$, and $\otimes_{k}$ denotes tensoring at position $k$. For such a Markovian master equation description to be valid, the system-bath couplings $\gamma_k$ must be small relative to the system energy scale $\Delta_k^{(l)}$. In addition, each dissipator acts only on the corresponding qutrit, i.e.~they are local. This requires that the strength of the interaction between the qutrits is at most comparable to the bath couplings $\gamma_k$ \cite{Hofer2017, Gonzalez2017}. We note that the reset model, while simple, can be mapped to a standard Lindblad-type model which can be derived from a microscopic, physical model of the baths \cite{Tavakoli2018,HaackInprep}.

The goal of the machine is to produce the $N$-qubit target state by local filtering of the $N$-qutrit steady state of \eqref{mastereq}. The steady state $\rho_{\infty}$ is obtained by solving $d\rho/dt=0$, and the filter is defined by a local projection $\Pi_k = \openone - \ket{R_k}\bra{R_k}$ of each qutrit onto the chosen qubit subspace. The state of the machine after filtering and the probability for the filtering to succeed are given by
\begin{align}\label{rhoout}
& \rho'=\frac{\Pi \rho_{\infty}\Pi}{\Tr\left(\rho_{\infty}\Pi\right)} && p_\text{suc}=\Tr\left(\rho_{\infty} \Pi\right),
\end{align}
where $\Pi = \bigotimes_{k=1}^N \Pi_k$. The temperatures, filters, bath couplings $\gamma_k$, and the interaction must be chosen appropriately for the heralded state $\rho'$ to approach the target state. 

Here, for a given $N$-qubit target $\ket{\psi}$, we focus on the following choice for the interaction
\begin{equation}\label{Hint}
H_{\text{int}}=g\left(\ketbra{\bar{\psi}}{R}+\ketbra{R}{\bar{\psi}}\right) ,
\end{equation}
where $g>0$ is the interaction strength, and the states $\ket{\bar{\psi}}$ and $\ket{R}$ are defined by the choices of filtered qubit subspace for each qutrit. For qutrit $k$, we let $R_k=0,1,2$ label the level which is \textit{not} part of the qubit, i.e.~qubit $k$ is spanned by the two levels complementary to $\ket{R_k}$. Then $\ket{\bar{\psi}}$ is the embedding of the target $\ket{\psi}$ into these qubit subspaces, and $\ket{R} = \ket{R_1\ldots R_N}$. That is, $H_{\text{int}}$ swaps the target state and the state in which every qutrit is outside the filtered subspace. For example, for $N=2$, if the target state is the maximally entangled two-qubit state $\ket{\psi}=\left(\ket{01}+\ket{10}\right)/\sqrt{2}$, and we choose $\ket{R}=\ket{20}$, then the embedding into the qutrits reads $\ket{\bar{\psi}}=\left(\ket{02}+\ket{11}\right)/\sqrt{2}$.

We furhermore focus on the regime of weak inter-system coupling, where $g$ is small relative to the free energies $\Delta_k^{(l)}$ (where the local master equation is valid). For there to be any non-trivial evolution in this regime, the interaction needs to be energy conserving, i.e.~$[H_{\text{int}},H_{\text{free}}]=0$. This restricts which target states can be generated. However, that is the only restriction. Our main result is that 

\begin{quote}
\textit{Any state $\ket{\psi}$, for which the Hamiltonians $H_{\text{free}}$ and $H_{\text{int}}$ of Eqs.~\eqref{Hfree} and \eqref{Hint} can be constructed to satisfy $[H_{\text{int}},H_{\text{free}}]=0$, can be generated by an entanglement engine as described above.}
\end{quote}

\noindent Specifically, one may choose a single qutrit to be connected with coupling strength $\gamma_h$ to a hot bath at temperature $T_h$ and all other qubits to be connected with coupling strength $\gamma_c$ to cold baths at $T_c$. For the hot qutrit, one chooses $R_k=2$, while for all the cold qutrits $R_k=0$. The target $\ket{\psi}$ is then obtained in the limit of extremal temperatures $T_c=0$, $T_h\rightarrow\infty$, and small coupling-strength ratios $g \lesssim \gamma_h \ll \gamma_c$. A full proof is given in \appref{MainProof}. However, one can intuitively understand why the machine works well in this regime. When $T_c=0$, resets of the cold qutrits will take them to the ground state $\ket{0}_k$. Since for the cold qutrits $R_k=0$, the ground state is not part of the filtered subspace. Therefore, cold resets will only lower the filtering success probability but will not affect the overlap of the filtered state with the target state $\ket{\psi}$. Once a cold qutrit is in the ground state, the only process which can bring it back into the filtered subspace is $H_{\text{int}}$, and this can only happen once all qutrits are in the state $\ket{R_k}$. The hot qutrit must then be in state $\ket{2}$, which can happen via a hot reset. Hot resets also degrade the quality of the filtered state (as they destroy coherence within the filtered subspace of the hot qutrit), and hence must be much less frequent than cold reset. This way, the system is most likely to be found outside the filtered subspace (making $p_{\text{suc}}$ small), but if found inside, it is likely to be in state $\ket{\psi}$ (because it is unlikely a hot reset happens before a cold one drives the system back out). The physical intuition for the bipartite case $N=2$ was also discussed in Ref.~\cite{Tavakoli2018}.

We note that, even if a given target $\ket{\psi}$ does not admit any choice of $H_{\text{free}}$ and $H_{\text{int}}$ satisfying $[H_{\text{int}},H_{\text{free}}]=0$, it may happen that by applying local unitaries to each qubit, one can obtain another state $\ket{\psi'}$ which does. Since entanglement is preserved under local unitaries, one may then first generate $\ket{\psi'}$ and simply apply the inverse local unitaries to obtain $\ket{\psi}$. Thus, effectively, the set of states which can be generated using the entanglement engine above consists of all states within the local unitary orbit of those $\ket{\psi}$ for which energy conservation can be satisfied.

\textit{Energy conservation.---}We now derive conditions for $\ket{\psi}$ to admit choices of $H_{\text{free}}$ and $H_{\text{int}}$ such that $\left[H_{\text{int}},H_{\text{free}}\right]=0$. This holds if and only if every transition generated by $H_{\text{int}}$ is energy conserving w.r.t.~$H_{\text{free}}$. From \eqref{Hint}, these transitions depend on the target state and on the choice of $\ket{R}$  (which defines the filtered qubit subspaces). We can write the target $N$-qubit state as
\begin{equation}
\label{target}
\ket{\psi} = \sum_{\mathbf{n} \in S_\psi} c_\mathbf{n} \ket{\mathbf{n}},
\end{equation}
where $S_\psi = \{\mathbf{n} \in \{0,1\}^N \, | \, \langle\psi | \mathbf{n}\rangle \neq 0 \}$ determines the set of basis states on which $\ket{\psi}$ has support, and $c_\mathbf{n} \in \mathbb{C}$. Denoting the embedding of $\ket{\mathbf{n}}$ into the $N$ qutrits 
%filtered qubit subspace
by $\ket{\bar{\mathbf{n}}}$, both $\ket{\bar{\mathbf{n}}}$ and $\ket{R}$ are eigenstates of $H_{\text{free}}$ with respective eigenvalues $E_{\mathbf{\bar{n}}}$ and $E_{R}$. The conditions for energy conservation are then $E_{\mathbf{\bar{n}}} = E_{R}$ for every $\mathbf{n} \in S_\psi$. This can be expressed as
\begin{align}
\label{autocond}
\frac{1}{2} \sum_{k=1}^N \bigg[ R_k n_k \Delta^{(1)}_k & + (2-R_k)((1-n_k)\Delta^{(1)}_k  + n_k \Delta^{(2)}_k) \bigg] \nonumber \\
&  - \frac{1}{2} \sum_{k=1}^N \bigg[ R_k \Delta^{(2)}_k \bigg] = 0 ,
\end{align}
where we have restricted to cases where the qubit states are either $\{\ket{1}_k,\ket{2}_k\}$ or $\{\ket{0}_k,\ket{1}_k\}$ for each qutrit (i.e.~$R_k=0$ or $R_k=2$) \footnote{Thermal resets on a given qutrit destroys entanglement with the other qutrits. For cold baths, thermal resets tend to drive the corresponding qutrit to the ground state. To suppress the effect of reset, it is therefore beneficial to choose $R_k=0$ when the bath temperature is cold. For infinitely hot baths, resets equalise the populations on the three levels, and it thus does not matter which subspace is filtered.}. Given a target state $\ket{\psi}$, the question is thus, whether there exist choices of $R_k$, $\Delta^{(1)}_k$, and $\Delta^{(2)}_k$ which fulfill \eqref{autocond} for all $\mathbf{n}\in S_\psi$.

Although \eqref{autocond} depends only on $S_\psi$ and not on the coefficients $c_\mathbf{n}$ in \eqref{target}, a general solution is not easy to obtain, because the number of variables increases with $N$. Nevertheless, \eqref{autocond} can be significantly simplified. In \appref{app.energycons}, we show that whenever \eqref{autocond} has a solution, then it has a solution with $R_k=0$ for all but a single $k$. For a given $\ket{\psi}$ it is thus sufficient to check whether there exists choices of $k'\in \{1,\ldots,N\}$, $\Delta^{(1)}_k$, and $\Delta^{(2)}_k$ fulfilling
\begin{equation}
\label{autocond2}
n_{k'}\Delta^{(1)}_{k'} + \sum_{k\neq k'} \bigg[ (1-n_k)\Delta^{(1)}_k  + n_k \Delta^{(2)}_k \bigg] - \Delta^{(2)}_{k'} = 0 .
\end{equation}
If there does, then it follows from the proof in \appref{MainProof} that the machine defined by these choices, with bath $k'$ hot and all other baths cold, can generate states arbitrarily close to $\ket{\psi}$.

Below, we consider several families of genuine multipartite entangled states, important in quantum information processing, namely GHZ, Dicke and cluster states. We show that they admit solutions to \eqref{autocond2} and hence can be generated. Furthermore, we consider the tradeoff between heralding success probability and the quality of the generated states, as well as the effect of finite temperatures, and show that they can be robustly generated also away from the ideal limit of the entanglement engine.

\begin{figure}
	\centering
	\includegraphics[width=\columnwidth]{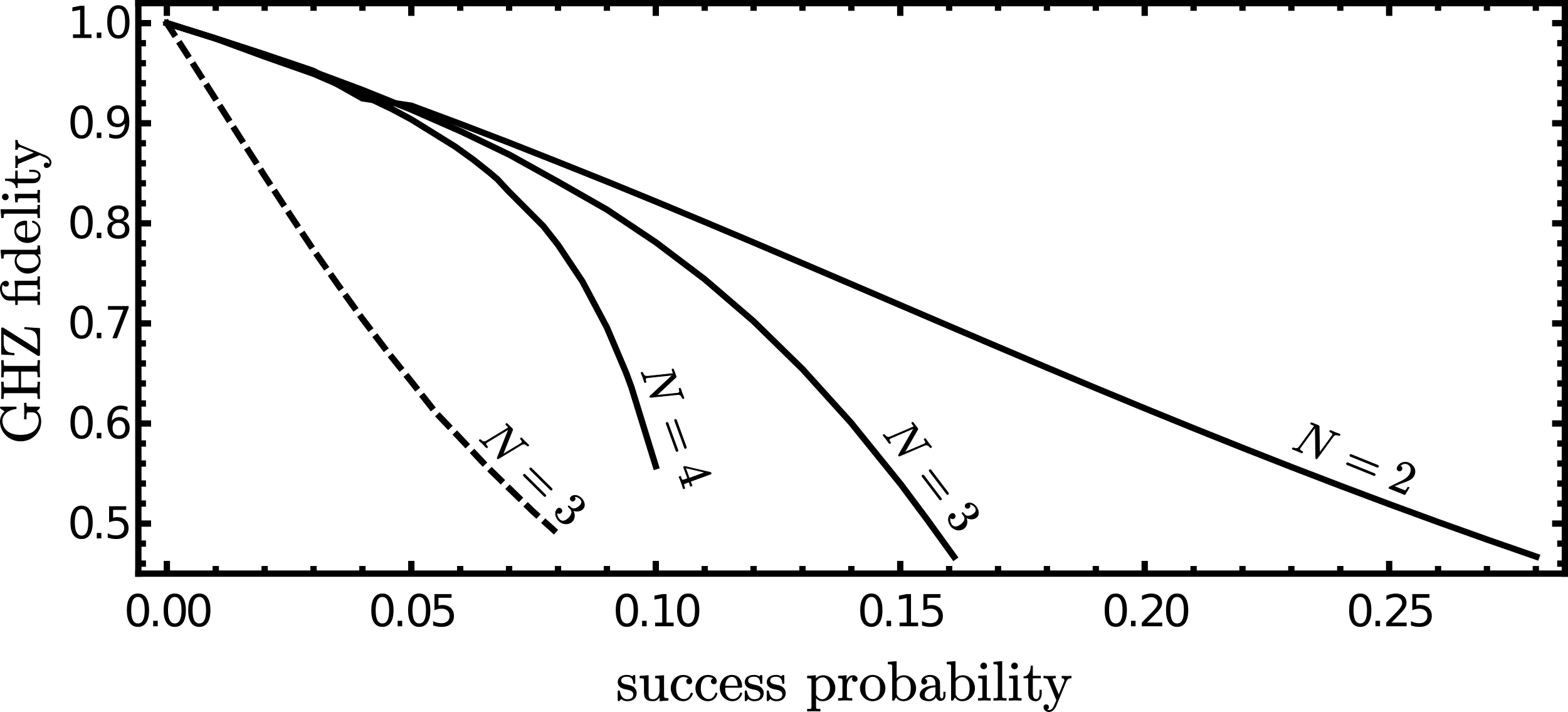}
	\caption{Fidelity of the generated state with the GHZ state versus the probability of successful filtering for different numbers of qutrits with one hot bath (solid lines) and two hot baths (dashed line). The curves are obtained by numerical optimisation over the coupling parameters under the constraint $g,\gamma_k\leq 10^{-2}\Delta_{\text{min}}$ where $\Delta_{\text{min}}$ is the smallest energy gap in each case.}\label{GHZfidelityFig}
\end{figure}

\textit{GHZ states.---}We start with the GHZ state of $N$ qubits, which is commonly given as $\frac{1}{\sqrt{2}}\left( \ket{0\ldots 0}+\ket{1\ldots 1} \right)$. In this form, the state does \textit{not} admit a solution to \eqref{autocond2}. However, we can instead consider $\ket{\text{GHZ}}=\frac{1}{\sqrt{2}} \left( \ket{10\ldots 0}+\ket{01\ldots 1} \right)$, which is equivalent up to a local unitary (bit flip) on the first party. One can check that $\ket{\text{GHZ}}$ does admit a solution to \eqref{autocond2}. One such solution is illustrated in \figref{fig.ghzmachine}. We take the first bath to be hot and the rest cold, and let the free Hamiltonians of the hot qutrit and each of the $N-1$ cold qutrits be 
\begin{align}
H_h & =  \Delta^{(1)}_h \ket{1}\bra{1} + \Delta^{(2)}_h \ket{2}\bra{2} , \\
H_c & =  \frac{\Delta^{(2)}_h-\Delta^{(1)}_h}{N-1} \ket{1}\bra{1} + \frac{\Delta^{(2)}_h}{N-1} \ket{2}\bra{2} .
\end{align}
To construct an energy-conserving interaction Hamiltonian, we follow the recipe above. Writing $\bar{0}$ for a string of $N-1$ zeros $0\ldots 0$, and similarly for $\bar{1}$ and $\bar{2}$, we have $\ket{R}=\ket{2 \bar{0}}$. Embedding $\ket{\text{GHZ}}$ in the qutrit space, from \eqref{Hint} we get
\begin{align}
H_{\text{int}} = g ( \ket{2 \bar{0}}\bra{1\bar{1}} + \ket{2\bar{0}}\bra{0\bar{2}} + \ket{1\bar{1}}\bra{2 \bar{0}} + \ket{0\bar{2}}\bra{2\bar{0}} ) ,
\end{align}
Once the steady state of the dynamics \eqref{mastereq} is obtained, we apply the filter $\Pi_h=\ketbra{0}{0}+\ketbra{1}{1}$ to the hot system and the filter $\Pi_c=\ketbra{1}{1}+\ketbra{2}{2}$ to each of the cold systems. Successful filtering heralds the generation of $\ket{\text{GHZ}}$.

As explained above, the perfect GHZ state is obtained only under idealised conditions (when the temperature gradient is maximal and the coupling strength ratios tend to zero). We now consider the quality of the generated state in case of finite temperatures and varying filtering success probabilities \eqref{rhoout}. We begin with the latter.

As argued above, in the ideal limit, $\gamma_h \ll \gamma_c$, the system is most likely found outside the filtered subspace, causing $p_{\text{suc}}\rightarrow 0$ as $\gamma_h/\gamma_c\rightarrow 0$. However, away from this idealised limit, we find that the state $\rho'$ after filtering (considered as an $N$-qubit state) may still have a high fidelity $F=\bracket{\text{GHZ}}{\rho'}{\text{GHZ}}$ with the GHZ state. \figref{GHZfidelityFig} shows the trade-off between $F$ and $p_{\text{suc}}$ for $N=2,3,4$ systems.  We see that fidelities above 90\% are obtained for $p_{\text{suc}}$ at the 5\%-level. Note that $p_{\text{suc}}$ is bounded, even when the fidelity is allowed to degrade. The maximal $p_{\text{suc}}$ decreases with increasing $N$, however the corresponding fidelity also increases. E.g.~for $N=4$, the fidelity does not reach $F=1/2$ before $p_{\text{suc}}$ reaches its maximal value of $p_{\text{suc}}=1/9$. This suggests that as $N$ grows, the fidelity achievable up to the maximal $p_{\text{suc}}$ increases. In \appref{GHZprob}, we derive the maximal value of $p_{\text{suc}}$ for any $N$. Finally, we note that we have also considered an analogous autonomous entanglement engine for $N=3$ with two hot systems and one cold system. However, as seen from \figref{GHZfidelityFig}, the performance in this case is worse.

\begin{figure}
	\centering
	\includegraphics[width=\columnwidth]{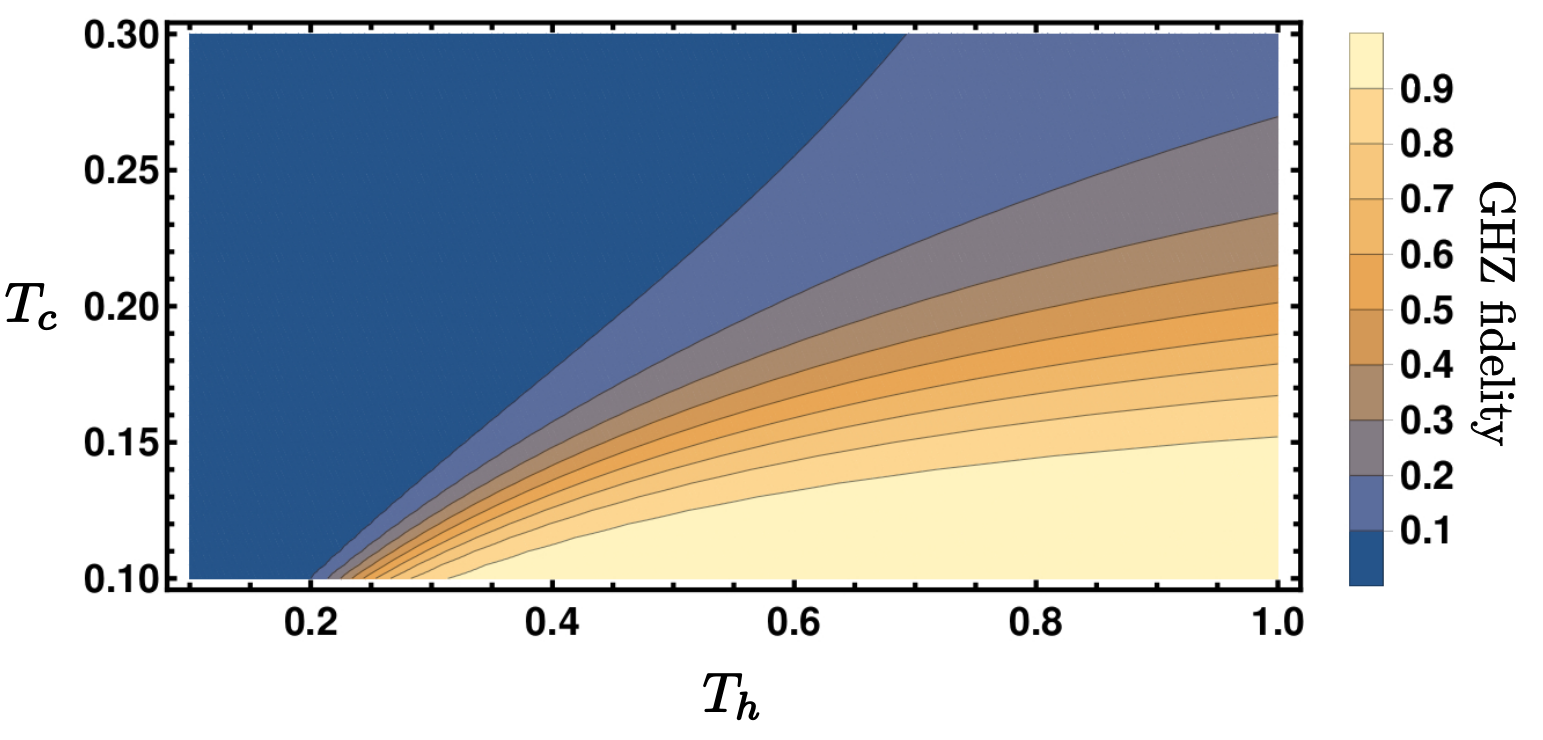}
	\caption{Fidelity of the filtered state with the GHZ state versus the bath temperatures, for $N=3$ and $g=1.6\times 10^{-3}$, $\gamma_h = 10^{-4}$, $\gamma_c=5\cdot 10^{-3}$, $\Delta^{(1)}=1$ $\Delta^{(2)}=2.5$.}\label{fig.GHZtemp}
\end{figure}

We remark that, for the states considered here which have only two non-zero off-diagonal elements, a GHZ fidelity $F > 1/2$ implies genuinely multipartite entanglement \cite{Guehne2010}. In addition, the $F>1/2$ also provides a certificate that this genuinely multipartite entanglement is strong enough to be semi-device-independently certified via the scheme of Ref.~\cite{Tavakoli2018b}. Furthermore, in \appref{app.GHZClusternonlocality} we have studied when the generated state can lead to Bell inequality violation (providing a fully device-independent certificate of entanglement). 

Next, we consider the effect of finite temperatures, i.e. $T_c > 0$ and $T_h < \infty$. We keep the interaction and bath coupling strengths fixed (thus also avoiding the idealised limit of vanishing couplings). The results are presented in \figref{fig.GHZtemp}. We note that even for temperatures far from the ideal limit, fidelities close to unity are possible. 

Thus, our entanglement engine functions well not only in the ideal limit but also for finite temperatures and coupling strengths. In \appref{app.lindblad}, we further show that qualitatively similar results can be obtained when the simple reset model is replaced by a master equation on standard Lindblad form, which can be derived from explicit, physical modeling of the baths and interactions.

\textit{Dicke states.---}As a second example, we consider $N$-qubit Dicke states. The Dicke state with $l$ excitations is given by 
\begin{equation}
\ket{D_l^N}=\frac{1}{\sqrt{\binom{N}{l}}}\sum_{s} \sigma_s\left[\ket{1}^{l}\otimes \ket{0}^{N-l}\right],
\end{equation} 
where the sum is over all permutations $\sigma_s$ of the subsystems. Notably, setting $l=1$ returns the well-known W-states. 

\begin{figure}[t]
	\centering
	\includegraphics[width=\columnwidth]{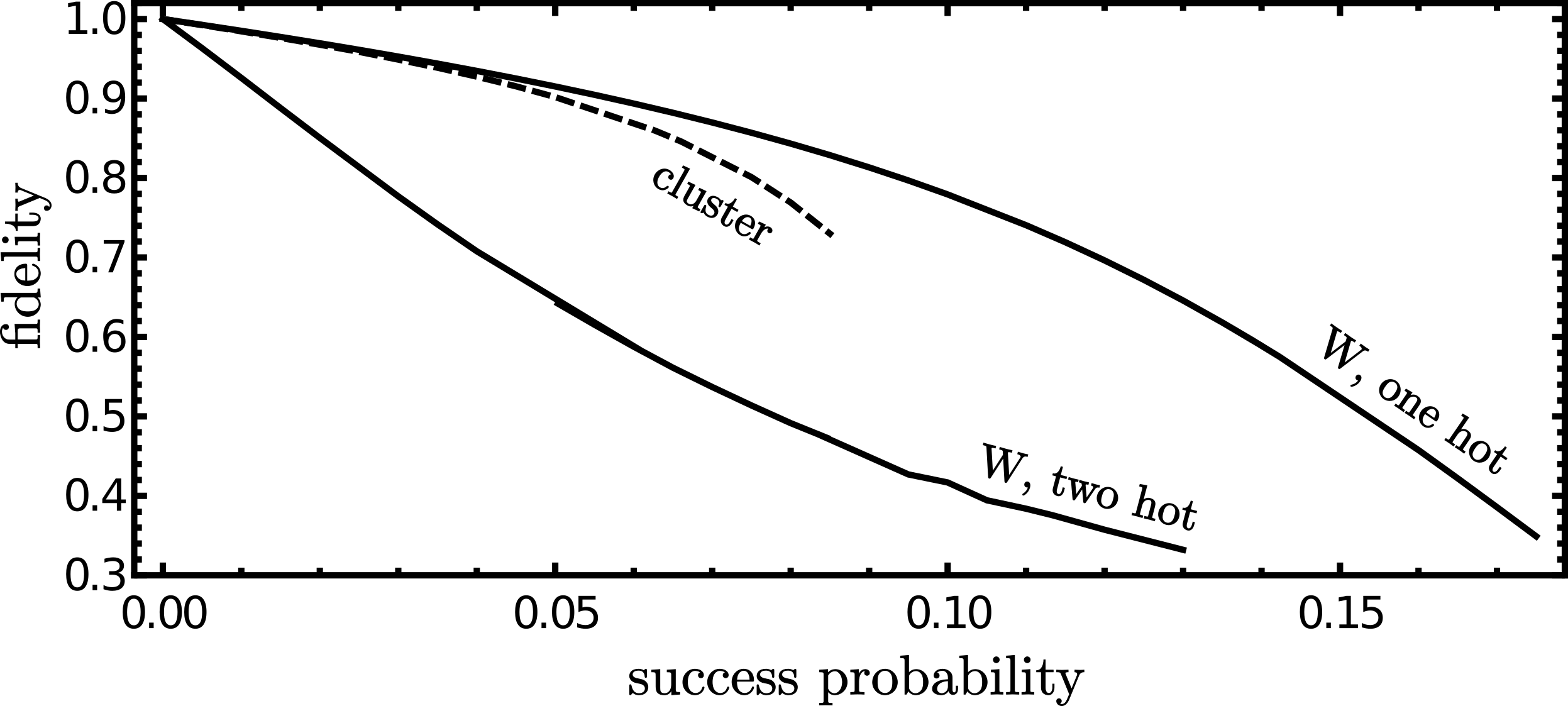}
	\caption{Fidelity versus the filtering success probability for generation of W-states using one and two hot baths (solid) and cluster states using one hot bath (dashed). The results are obtained by constrained optimisation over $\gamma_h,\gamma_c,g\leq 10^{-2}\Delta_{\text{min}}$, where $\Delta_{\text{min}}$ is the smallest energy gap in each case.}\label{WCfig}
\end{figure}

Again, one finds that all such states admit solutions to \eqref{autocond2}. Hence, every Dicke state can be generated by an autonomous entanglement engine. For instance, we choose the first qutrit hot and the rest cold, and the free Hamiltonians
$H_h =  \Delta^{(1)}_h \ket{1}\bra{1} + \Delta^{(2)}_h \ket{2}\bra{2}$ and $H_c =  \Delta^{(1)}_c \ket{1}\bra{1} + \Delta^{(2)}_c \ket{2}\bra{2}$, where
\begin{align}
\Delta_h^{(1)} & = \left(N-1+(l-1)\left(\Delta^{(2)}_c-\Delta^{(1)}_c\right)\right) , \\
\Delta_h^{(2)} & = N-1+l\left(\Delta^{(2)}_c-\Delta^{(1)}_c\right) .
\end{align}
Note that similar solutions of \eqref{autocond} are possible also for more hot baths. For the case $(N,l)=(3,1)$, we have analytically solved the reset master equation in terms of $g,\gamma_h,\gamma_c$ and computed the fidelity $F=\bracket{D_1^3}{\rho'}{D_1^3}$. Similarly, we have analytically evaluated $p_\text{suc}$ in \eqref{rhoout}. The tradeoff between $F$ and $p_{\text{suc}}$ is shown in \figref{WCfig}. As for the GHZ case, we find that high fidelities can be reached with success probabilities at the few-percent level. We have also checked that increasing the number of hot systems (to two) does not improve performance.

\textit{Cluster state.---}Finally, we consider a linear four-qubit cluster state
\begin{multline}
\label{eq.cluster}
\ket{C}=\frac{1}{2}\big(\ket{0110}+\ket{0101}+\ket{1010}-\ket{1001}\big).
\end{multline}
A solution to \eqref{autocond2} is obtained by the following free Hamiltonian, where $\Delta = \Delta^{(2)}_c-\Delta^{(1)}_c$
\begin{align}
H_h & =  \left(3+\Delta\right)\ket{1}\bra{1} + \left(3+2\Delta\right) \ket{2}\bra{2},\\
H_c & = \Delta^{(1)}_c \ket{1}\bra{1} +\Delta^{(2)}_c\ket{2}\bra{2}.
\end{align}

In analogy with the previous, we consider the trade-off between the $F=\bracket{C}{\rho'}{C}$ of the generated state $\rho'$ with the cluster state and filtering success probability $p_\text{suc}$. We have evaluated both $F$ and $p_\text{suc}$ analytically for a single hot bath, and optimised over the couplings $g,\gamma_h,\gamma_c$ to obtain the results in \figref{WCfig}. Again, high-fidelity cluster states can be generated with success probabilities at the few-percent level. Furthermore, in \appref{app.GHZClusternonlocality}, we have considered the device-independent certification of $\rho'$ via Bell inequalities tailored for cluster states \cite{Scarani2005} at varying $p_\text{suc}$. We find that large Bell inequality violations can be obtained for every $p_\text{suc}$ up to its maximal value of $p_{\text{suc}} \approx 0.085$, demonstrating that the entanglement engine works well over a wide regime.

\textit{Conclusion.---}We have given a general recipe for autonomous entanglement engines which enable heralded generation of multipartite entangled states between any number of parties. As demonstrated by several examples, a wide range of states can be targeted, including GHZ, Dicke, and cluster states. While pure target states are only generated perfectly for infinite temperature gradients and vanishing heralding success probabilities, we have explored finite temperatures and heralding probabilities as well and have found that high fidelities can be attained also away from the ideal regime. 

Thus, probabilistic generation of high-quality multipartite entanglement is possible using only incoherent, thermal processes and energy-preserving interactions, requiring no work input. It would be interesting to understand if strong entanglement could be generated by an autonomous engine in a deterministic manner, i.e.~without filtering. Finally, perspectives for experimental implementation could be explored. In that context, a natural question is whether genuine multipartite entangled states can be generated autonomously using only two-body Hamiltonians.

\textit{Acknowledgements.---}We thank Marcus Huber for discussions. JBB was supported by the Independent Research Fund Denmark, AT and NB by the Swiss National Science Foundation (Grant 200021\_169002 and NCCR QSIT), and GH by the Swiss National Foundation through the starting grant PRIMA PR00P2$\_$179748 .

\bibliography{multipartite_thermalent.bib}

%%%%%%%%%%%%%%%%%%%%%%%%%%%%%%%%%%%%%%%%%%%%%%%%%%%%%%%%%%%%%%%%%%%
% Appendices
%%%%%%%%%%%%%%%%%%%%%%%%%%%%%%%%%%%%%%%%%%%%%%%%%%%%%%%%%%%%%%%%%%%

\onecolumngrid
\appendix

\section{Autonomous generation of target states}
\label{MainProof}

We prove that any state $\ket{\psi}$ which admits a solution to the energy-conservation condition \eqref{autocond} can be generated by an autonomous entanglement engine. Following the main text, we write the target state as 
\begin{equation}
\ket{\psi}=\sum_{\mathbf{n}\in S_\psi} c_\mathbf{n}\ket{\mathbf{n}}
\end{equation}
where $c_\mathbf{n}\in\mathbb{C}$, $\sum_\mathbf{n}|c_\mathbf{n}|^2=1$, and where $S_\psi$ is the set of binary strings $s=\{0,1\}^N$ such that $\ket{\psi}$ has support of $\ket{s}$. We show that the state $\rho'$ returned by the machine described in the main text (after heralding) is indeed the target state. To this end, we must characterise $\rho'$. For simplicity, we will first focus on the diagonal elements of $\rho'$ and then on its off-diagonal elements.

\subsection{Diagonal elements}

\begin{figure}[t]
	\begin{center}
		\includegraphics[width=0.45\linewidth]{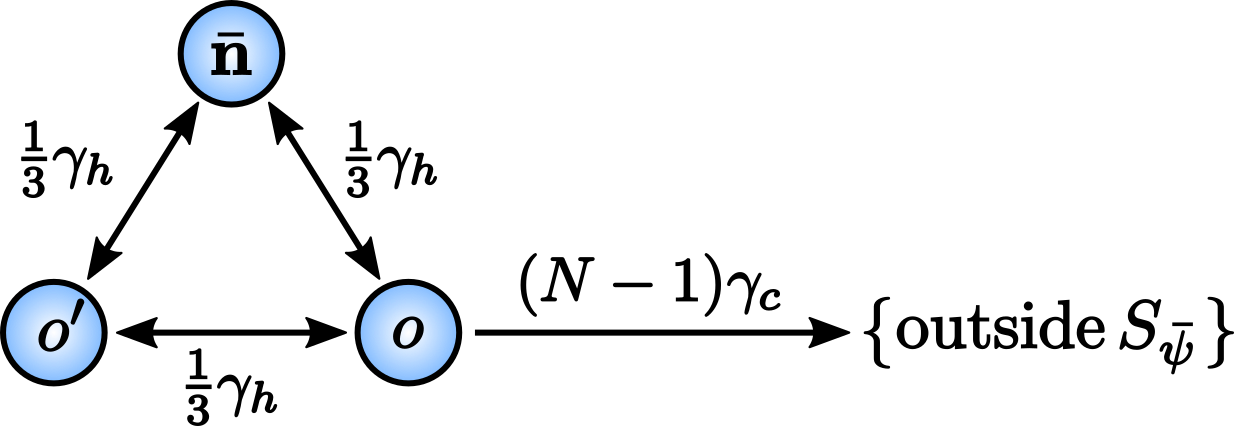}
		\caption{Flow diagram for population entering and leaving the state $\ket{o}$. Hot resets take the system from $\ket{o}$ to states $\ket{ö́'}$ or $\ket{\bar{\mathbf{n}}}$, while cold resets take it to other states outside the support $S_{\bar{\psi}}$. The transition rates due to hot and cold resets are indicated.}
		\label{fig.diagelflow}
	\end{center}
\end{figure}

We aim to show that the diagonal elements of $\rho'$ correspond to the populations $|c_{\bar{\mathbf{n}}}|^2$, where $\ket{\bar{\mathbf{n}}}$ are the computational basis states on which the embedded target state $\ket{\bar{\psi}}$ has support.  To enable the characterisation of the diagonal elements of  $\rho'$, we use flow diagrams as illustrated in \figref{fig.diagelflow}. Such a diagram represents the transitions induced by the influence of hot and cold resets, along with the rate of said transitions, on a given support state $\ket{\bar{\mathbf{n}}}$. As illustrated; by a hot reset on $\ket{\bar{\mathbf{n}}}$ one can reach two other states, denoted by $\ket{o}$ and $\ket{o'}$. Importantly, neither of these two states can be members of $S_{\bar{\psi}}$ since it is otherwise at odds with the conditions for an autonomous Hamiltonian. From the flow-diagram, we obtain the following steady-state condition when considering the flow into and out of the state $\ket{o}$:
\begin{equation}\label{flowsteady}
P_{o}\left(2\frac{\gamma_h}{3}+\gamma_c(N-1)\right)=\frac{\gamma_h}{3}\left(P_{\bar{\mathbf{n}}}+P_{o'}\right),
\end{equation} 
where we have adopted the simplified notation $P_s=\bracket{s}{\rho}{s}$. However, since $\ket{o},\ket{o'}\notin S_{\bar{\psi}}$ (nor do they equal the state $\ket{R}$), they do not appear in the interaction Hamiltonian and are treated equally by the dissipation. Hence, it follows that $P_{o}=P_{o'}$. This leads us to re-write \eqref{flowsteady} as
\begin{equation}\label{ratio}
\frac{P_o}{P_{\bar{\mathbf{n}}}}=\frac{\gamma_h}{3(N-1)\gamma_c+\gamma_h}.
\end{equation}

Let us now consider the filtered subspace, i.e. the space in which the heralded state $\rho'$ lives. Since the filtering corresponds to projecting each qutrit onto a qubit subspace, there are consequently $2^N$ computational basis states spanning the filtered sub-space. Of these, $\nu= |S_{\bar{\psi}}|$ are members of $S_{\bar{\psi}}$, whereas another $\nu$ are reachable by a hot reset to each element in $S_{\bar{\psi}}$. Denote the latter set of states by $G_h$. The remaining $2^N-2\nu$ states have no population (diagonal element equal zero) since they can neither be reached via the interaction Hamiltonian nor via resets. Let $\bar{P}_o$ denote renormalised $P_o$ after filtering, i.e., $\bar{P}_o=\bracket{o}{\rho'}{o}$. Normalisation requires that
\begin{equation}
\sum_{o\in S_{\bar{\psi}}} \bar{P}_o+\sum_{o\in G_h}\bar{P}_o=1.
\end{equation}
However, due to the symmetries of the interaction Hamiltonian and the linearity of the dynamics, we may write $\bar{P}_{o}=|c_{o}|^2 \bar{P}_{\text{S}}$ for $o\in S_{\bar{\psi}}$ for some  constant population $\bar{P}_{\text{S}}$ independent of $o$. Similarly, we may write $\bar{P}_{o}=|c_{o}|^2\bar{P}_{\text{G}}$ for $o\in G_h$ for some constant population $\bar{P}_{\text{G}}$ independent of $o$. The normalisation condition reduces to 
\begin{equation}
\bar{P}_{\text{S}}\left(1+\frac{\bar{P}_{\text{G}}}{\bar{P}_{\text{S}}}\right)=1
\end{equation}
which together with \eqref{ratio} gives
\begin{align}
& \bar{P}_{\text{S}}=\left(1+\frac{\bar{P}_{\text{G}}}{\bar{P}_{\text{S}}}\right)=\left(1+\frac{\gamma_h}{3(N-1)\gamma_c+\gamma_h}\right)^{-1}.
\end{align}
In the limit $\gamma_h\ll \gamma_c$ we have $\bar{P}_{\text{S}}\rightarrow 1$, and therefore also $\bar{P}_{\text{G}}\rightarrow 0$. Consequently, we have found that in the given limit, for $\bar{\mathbf{n}}\in S_{\bar{\psi}}$
\begin{equation}\label{populations}
\bar{P}_{\bar{\mathbf{n}}}=\bracket{\bar{\mathbf{n}}}{\rho'}{\bar{\mathbf{n}}}=|c_{\bar{\mathbf{n}}}|^2. 
\end{equation}
These are the desired diagonal elements.

\subsection{Off-diagonal elements}
We now aim to show that the off-diagonal elements of $\rho'$ correspond to $c_nc_n^*$. Due to hermiticity, it is sufficient to consider the upper triangle in the matrix of $\rho'$. Among these off-diagonal entries, there are $\binom{\nu}{2}$ that correspond to coherences generated between the computational basis states associated to $n,n'\in S_{\bar{\psi}}$ (we have dropped the notation in bold ($\bar{\mathbf{n}}$) since in this section $n$ will sometimes be a member of $S_{\bar{\psi}}$). Another $\nu$ off-diagonals correspond to coherences generated between the computational basis states assciated to $n\in S_{\bar{\psi}}$ and the state $\ket{R}$. The remaining off-diagonal elements are not reachable by the dynamics (neither via resets nor via the Hamiltonian) and therefore equal zero. We use the short-hand notation $\rho_{n,n'}=\bracket{n}{\rho}{n'}$ to write the reset master equation in the steady state as
\begin{equation}\label{masterproof}
0=\dot{\rho}_{n,n'}=-i\bracket{n}{[H,\rho]}{n'}+\frac{\gamma_h}{3}\bracket{n}{\openone\otimes \Tr_1\left(\rho\right)}{n'}
+\sum_{k=2}^{N}\gamma_c\bracket{n}{\left(\ketbra{0}{0}\otimes_k \Tr_k\left(\rho\right)\right)}{n'}-\left(\gamma_h+\gamma_c\right)\rho_{n,n'}.
\end{equation}

For the first term in Eq.~\eqref{masterproof} we have that
\begin{multline}
\bracket{n}{[H,\rho]}{n'}=g\bracket{n}{\left(\ketbra{\bar{\psi}}{R}+\ketbra{R}{\bar{\psi}}\right)\rho-\rho\left(\ketbra{\bar{\psi}}{R}+\ketbra{R}{\bar{\psi}}\right)}{n'}\\
=g\left(\braket{n}{\bar{\psi}}\bracket{R}{\rho}{n'}+\braket{n}{R}\bracket{\bar{\psi}}{\rho}{n'}-\bracket{n}{\rho}{\bar{\psi}}\braket{R}{n'}-\bracket{n}{\rho}{R}\braket{\bar{\psi}}{n'}\right).
\end{multline}
Taking $n,n'\neq R$, the two middle terms vanish. Moreover, if $n,n'\notin S_{\bar{\psi}}$  also the first and fourth term vanish. If $n,n'\in S_{\bar{\psi}}$ then we have $\braket{n}{\bar{\psi}}=c_n$ and $\braket{\bar{\psi}}{n'}=c_{n'}^*$ and therefore $\bracket{n}{[H,\rho]}{n'}=g\left(c_n\rho_{R,n'}-c_{n'}^*\rho_{n,R}\right)$. Thus,
\begin{equation}
\bracket{n}{[H,\rho]}{n'}=\begin{cases}
g\left(c_n \rho_{R,n'}-c_{n'}^*\rho_{n,R}\right) & \text{if } n,n'\in S_{\bar{\psi}}\\
0              & \text{if } n,n'\notin S_{\bar{\psi}} \text{ and } n,n'\neq R  
\end{cases}.
\end{equation}
For the second term in Eq.~\eqref{masterproof} a direct calculation gives
\begin{equation}\label{sec}
\bracket{n}{\openone\otimes \Tr_1\left(\rho\right)}{n'}=\delta_{n_1,n_1'}\sum_j \rho_{j\bar{n},j\bar{n}'},
\end{equation}
where the bar-sign denotes $\bar{s}=s_2\ldots s_N$. Moreover, the third term in \eqref{masterproof} straightforwardly evaluates to
\begin{equation}
\bracket{n}{\left(\ketbra{0}{0}\otimes_k \Tr_k\left(\rho\right)\right)}{n'}=\delta_{n_k,0}\delta_{n_k',0}\sum_{j_k}\rho_{\overleftarrow{n}j_k\overrightarrow{n},\overleftarrow{n}'j_k\overrightarrow{n}'},
\end{equation}
where $\overleftarrow{s}=s_1\ldots s_{k-1}$ and $\overrightarrow{s}=s_{k+1}\ldots s_N$. Notice that this term vanishes for $k=2,\ldots,N$ if either $n$ or $n'$ are members of $S_{\bar{\psi}}$. In conclusion, for $n,n'\in S_{\bar{\psi}}$, we can re-write \eqref{masterproof} as
\begin{equation}\label{masterproof2}
0=\dot{\rho}_{n,n'}=-ig\left(c_n \rho_{R,n'}-c_{n'}^*\rho_{n,R}\right)+\frac{\gamma_h}{3}\delta_{n_1,n_1'}\sum_j \rho_{j\bar{n},j\bar{n}'}-\left(\gamma_h+\gamma_c\right)\rho_{n,n'}.
\end{equation}

When $n\neq n'$ (since one cannot transition between two support states by a hot reset) Eq.~\eqref{sec} becomes  $\delta_{n_1,n_1'}\sum_j \rho_{j\bar{n},j\bar{n}'}=\delta_{n_1,n_1'}\rho_{n,n'}$. Furthermore, by hermiticity we have that $\rho_{R,n'}=\rho_{n',R}^*$, and due to the symmetries of the Hamiltonian it also holds that $\rho_{n,R}=c_nL$ where $L$ is a constant related to the population in the steady-state that is independent of $n$. With this in hand, we consider the three equations obtained from \eqref{masterproof2}:
\begin{eqnarray}\label{eq1}
0=\dot{\rho}_{n,n'}=-igc_nc_{n'}^*\left(L^*-L\right)+\frac{\gamma_h}{3}\delta_{n_1,n_1'}\rho_{n,n'}-\left(\gamma_h+\gamma_c\right)\rho_{n,n'}\\\label{eq2}
0=\dot{\rho}_{n,n}=-ig|c_n|^2\left(L^*-L\right)+\frac{\gamma_h}{3}\sum_j \rho_{j\bar{n},j\bar{n}}-\left(\gamma_h+\gamma_c\right)\rho_{n,n}\\\label{eq3}
0=\dot{\rho}_{i\bar{n},i\bar{n}}=\frac{\gamma_h}{3}\sum_j \rho_{j\bar{n},j\bar{n}}-\left(\gamma_h+\gamma_c\right)\rho_{i\bar{n},i\bar{n}},
\end{eqnarray}
where in the first equation we have taken $n,n'\in S_{\bar{\psi}}$ with $n\neq n'$, in the second equation we have taken $n,n'\in S_{\bar{\psi}}$ with $n=n'$, and in the third equation we have taken $n,n'\in S_{\bar{\psi}}$ with $n=n'$ but then replaced $n_1$ with the index $i$ which runs over the two values $i\neq n_1$. Summing over $i$ in the  equation \eqref{eq3} gives
\begin{equation}
\sum_{i\neq n_1}\rho_{i\bar{n},i\bar{n}}=\frac{2\gamma_h}{3\gamma_c+\gamma_h}\rho_{n,n}.
\end{equation}
Inserted into the equation \eqref{eq2} we obtain
\begin{equation}
ig(L^*-L)=-\frac{\rho_{n,n}}{|c_n|^2}\frac{3\gamma_c\left(\gamma_h+\gamma_c\right)}{3\gamma_c+\gamma_h}.
\end{equation}
Finally, when inserted into the equation \eqref{eq1}, we can obtain the off-diagonal elements from the diagonal elements of $\rho'$. We obtain
\begin{equation}
\rho_{n,n'}=-\frac{3\gamma_c\left(\gamma_h+\gamma_c\right)}{3\gamma_c+\gamma_h}\left(\frac{\gamma_h}{3}\delta_{n_1,n_1'}-\left(\gamma_h+\gamma_c\right)\right)^{-1}\frac{c_nc_{n'}^*}{|c_n|^2}\rho_{n,n}
\end{equation}
However, the ratios between the off-diagonal terms are conserved after filtering if they belong to the filtered subspace. We use the notation $\bar{\rho}_{s,s'}=\bracket{s}{\rho'}{s'}$. Then, taking the relevant limit of $\gamma_h\ll\gamma_c$, we obtain
\begin{equation}
\lim_{\gamma_h\ll \gamma_c}\bar{\rho}_{n,n'}=\frac{c_nc_{n'}^*}{|c_n|^2}\lim_{\gamma_h\ll\gamma_c}\bar{\rho}_{n,n}.
\end{equation}
The right-hand-side features a diagonal element which was evaluated in \eqref{populations}. In the relevant limit, we obtain the final result 
\begin{equation}\label{coherences}
\lim_{\gamma_h\ll \gamma_c}\bar{\rho}_{n,n'}=c_nc_{n'}^*.
\end{equation}
In conclusion, we have shown that the heralded state  $\rho'$ is the target state.

\section{Simplified conditions for energy conservation}
\label{app.energycons}

\subsection{A single hot system is sufficient}
\label{app.onehotproof}

Here, we show that if the conditions \eqref{autocond} for the interaction to be energy conserving can be solved using $q$ hot systems (i.e.~systems with $R_k=2$) and $N-q$ cold systems (i.e.~systems with $R_k=0$), then there also exists a solution with just a single hot system and $N-1$ cold systems.

To prove this, we show that any set of valid energies $\Delta_k^{(1)}$, $\Delta_k^{(2)}$ fulfilling the energy-conservation condition for $q$ hot systems allows one to define another set of energies $\{\varepsilon_k^{(1)}$, $\varepsilon_k^{(2)}\}$ which fulfill the corresponding condition with a single hot system. Without loss of generality (as one may always permute the parties), we can take the hot systems to be the first ones. Then the energy-conservation condition with $q$ hot systems reads
\begin{equation}\label{qhot}
\forall \mathbf{n}\in S_\psi: \quad\sum_{k=1}^{q}\left(n_k\Delta_k^{(1)}-\Delta_k^{(2)}\right)+\sum_{k=q+1}^{N}\left((1-n_k)\Delta_{k}^{(1)}+n_k\Delta_k^{(2)}\right)=0 ,
\end{equation}
while the corresponding condition with a single hot system ($q=1$) becomes
\begin{equation}\label{1hot}
\forall \mathbf{n}\in S_\psi: \quad \left(n_1\varepsilon_1^{(1)}-\varepsilon_1^{(2)}\right)+\sum_{k=2}^{N}\left((1-n_k)\varepsilon_{k}^{(1)}+n_k\varepsilon_k^{(2)}\right)=0 .
\end{equation}
Note that the energies must satisfy $\Delta_k^{(2)} > \Delta_k^{(1)} > 0$ and similarly $\varepsilon_k^{(2)} > \varepsilon_k^{(1)} > 0$. To construct a solution to \eqref{1hot} given a solution to \eqref{qhot}, we choose
\begin{align}
\begin{array}{l}
\varepsilon_k^{(1)} = \Delta_k^{(1)} \\[0.25cm] 
\varepsilon_k^{(2)} = \Delta_k^{(2)}
\end{array} \quad\quad \text{for} \quad k = q+1,\ldots,N , \\
\end{align}
and
\begin{align}
\begin{array}{l}
\varepsilon_k^{(1)}=t_k-\Delta_k^{(2)} \\[0.25cm] 
\varepsilon_k^{(2)}=t_k-\Delta_k^{(2)}+\Delta_k^{(1)}
\end{array} \quad\quad \text{for} \quad k = 2,\ldots,q , \\
\end{align}
for some $t_k$ satisfying $t_k>\Delta_k^{(2)}$. Note that with these choices we have $\varepsilon_k^{(2)} > \varepsilon_k^{(1)} > 0$ for $k=2,\ldots,N$, as desired. Inserting in \eqref{1hot}, we get
\begin{equation}\label{1hot2}
\forall \mathbf{n}\in S_\psi: \quad \left(n_1\varepsilon_1^{(1)}-\varepsilon_1^{(2)}+\sum_{k=2}^{q}t_k\right)+\sum_{k=2}^{q}\left(n_k\Delta_{k}^{(1)}-\Delta_{k}^{(2)}\right)+\sum_{k=q+1}^{N}\left((1-n_k)\Delta_{k}^{(1)}+n_k\Delta_k^{(2)}\right)=0.
\end{equation}
This reduces to \eqref{qhot} provided that 
\begin{equation}
\forall n_1: \quad \left(n_1\varepsilon_1^{(1)}-\varepsilon_1^{(2)}+\sum_{k=2}^{q}t_k\right)=n_1\Delta_{1}^{(1)}-\Delta_{1}^{(2)} ,
\end{equation}
which is solved by
\begin{align}
\varepsilon_1^{(1)} & = \Delta_1^{(1)} , \\
\varepsilon_1^{(2)} & = \Delta_1^{(2)} + \sum_{k=2}^q t_k .
\end{align}
It is easy to see that $\varepsilon_1^{(2)} > \varepsilon_1^{(1)} > 0$. We thus have a valid choice of energies $\varepsilon_k^{(1)}$, $\varepsilon_k^{(1)}$ for which \eqref{1hot} reduces \eqref{qhot}. Hence, any solution with $q$ hot systems also implies the existence of a solution with a single hot system, as claimed.

\subsection{Identical energy structures for all hot and all cold systems}
\label{app.identicalenergies}

If the energy spectra of all hot systems (i.e.~all systems with $R_k=2$) are identical, and similarly those of cold systems (with $R_k=0$) are identical, then the energy-conservation conditions can be simplified. Note that all the examples given in the main text (for GHZ, Dicke, and cluster states) belong to this setting. 

Specifically, here we show that if $\Delta^{(1)}_k$ and $\Delta^{(2)}_k$ depend only on $R_k$, then the existence of $R\in\{0,2\}^N$ and a choice of energies fulfilling \eqref{autocond} in the main text is equivalent to the existence of a vector $\mathbf{r} \in\{0,1\}^N$ such that $\mathbf{r} \neq \mathbf{0}, \mathbf{1}$ and for each pair of vectors $\mathbf{n}, \mathbf{n}' \in S_\psi$ either
\begin{equation}
\label{eq.conscond1}
\left(\mathbf{n}-\mathbf{n}'\right) \cdot \mathbf{r} = \left(\mathbf{n}-\mathbf{n}'\right) \cdot \left(\mathbf{1}-\mathbf{r}\right) = 0 ,
\end{equation}
or
\begin{equation}
\label{eq.conscond2}
\frac{\left(\mathbf{n}-\mathbf{n}'\right) \cdot \mathbf{r}}{ \left(\mathbf{n}-\mathbf{n}'\right) \cdot \left(\mathbf{1}-\mathbf{r}\right)} = c
\end{equation}
where $c<0$ is a constant independent of $\mathbf{n},\mathbf{n}'$, and $\mathbf{0}=(0,0,\ldots,0)$ and $\mathbf{1} = (1,1,\ldots,1)$. That is, the interaction \eqref{Hint} can be made energy conserving if and only if an $\mathbf{r} \neq \mathbf{0}, \mathbf{1}$ exists fulfilling \eqref{eq.conscond1}-\eqref{eq.conscond2}.

Before we proceed with the proof, we illustrate \eqref{eq.conscond1}-\eqref{eq.conscond2} and the notation introduced above in the simplest setting of two parties. We take the maximally entangled state $\ket{\Psi^+}=\left(\ket{01}+\ket{10}\right)/\sqrt{2}$ as the target and choose $\ket{R} = \ket{20}$. The target has support on just two states, $S_\psi = \{\mathbf{n},\mathbf{n}'\}$, where
\begin{equation}
\mathbf{n} = (0,1) \quad \text{and} \quad \mathbf{n}' = (1,0) .
\end{equation}
It is straightforward to verify that \eqref{eq.conscond2} is satisfied for $\mathbf{r} = (1,0)$, with $c=-1$. Hence, $\ket{\Psi^+}$ can indeed be generated autonomously. Looking at \eqref{autocond}, we see that the conditions on the energies coming from $\mathbf{n}$ and $\mathbf{n}'$ are respectively
\begin{equation}
\Delta^{(2)}_2 = \Delta^{(2)}_1 ,
\end{equation}
and
\begin{equation}
\Delta^{(1)}_2 = \Delta^{(2)}_1 - \Delta^{(1)}_1 .
\end{equation}
Thus, the two qutrits have the same maximal energy but inverted level structures. The gap between the two lower levels for the second qutrit equals the gap between the upper two levels for the first qutrit. This corresponds exactly to the entanglement engine of Ref.~\cite{Tavakoli2018}.

The conditions \eqref{eq.conscond1}-\eqref{eq.conscond2} can be defined as follows. If we define a vector $\mathbf{r}\in\{0,1\}^N$ such that $r_k=0$ if $R_k=0$ and $r_k=1$ for $R_k=2$, then for each $\mathbf{n} \in S_\psi$, the condition $E_{\mathbf{n}} = E_{\bar{R}}$ from \eqref{autocond} of the main text can be expressed as
\begin{align}
\sum_{k=1}^N \left[ r_k n_k \Delta^{(1)}_k + (1-r_k)((1-n_k)\Delta^{(1)}_k + n_k \Delta^{(2)}_k)  - r_k \Delta^{(2)}_k \right] = 0 .
\end{align}
The question is, whether there exist choices of $\mathbf{r}$, $\Delta^{(1)}_k$, and $\Delta^{(2)}_k$ which fulfill this. Rewriting, we have
\begin{align}
\label{eq.sumcond}
\sum_{k \,\text{s.t.}\, r_k=0} \left[ (1-n_k)\Delta^{(1)}_k + n_k \Delta^{(2)}_k \right] + \sum_{k \,\text{s.t.}\, r_k=1} \left[ n_k \Delta^{(1)}_k - \Delta^{(2)}_k \right] = 0 .
\end{align}
Now, if the energy structures of all qutrits with the same $R_k$ are the same, then the energies appearing under each sum become independent of $k$. Let us denote the energy gaps of qutrits with $R_k=0$ by $\delta_1=\Delta^{(1)}_k$ and $\delta_2=\Delta^{(2)}_k-\Delta^{(1)}_k$ and those of qutrits with $R_k=2$ by $\delta_3=\Delta^{(1)}_k$ and $\delta_4=\Delta^{(2)}_k-\Delta^{(1)}_k$. Then \eqref{eq.sumcond} becomes
\begin{equation}
\sum_{k \,\text{s.t.}\, r_k=0} \left[ \delta_1 + n_k \delta_2 \right] + \sum_{k \,\text{s.t.}\, r_k=1} \left[ (n_k-1) \delta_3 - \delta_4 \right] = 0 ,
\end{equation}
which is equivalent to
\begin{align}
(N-|\mathbf{r}|) \delta_1 + (\mathbf{1}-\mathbf{r})\cdot \mathbf{n} \, \delta_2 - \mathbf{r}\cdot(\mathbf{1}-\mathbf{n}) \delta_3 - |\mathbf{r}| \delta_4 = 0 ,
\end{align}
where $\mathbf{1} = (1,\ldots,1)$ and $|\mathbf{r}|$ is the number of 1's in $\mathbf{r}$. This must hold for every $\mathbf{n} \in S_\psi$, and thus we have a set of linear equations
\begin{equation}
\label{eq.lineqsys}
\begin{pmatrix}
N-|\mathbf{r}| & (\mathbf{1}-\mathbf{r})\cdot \mathbf{n}^{(1)} & - \mathbf{r}\cdot(\mathbf{1}-\mathbf{n}^{(1)}) & - |\mathbf{r}| \\
N-|\mathbf{r}| & (\mathbf{1}-\mathbf{r})\cdot \mathbf{n}^{(2)} & - \mathbf{r}\cdot(\mathbf{1}-\mathbf{n}^{(2)}) & - |\mathbf{r}| \\
\vdots & & \\
N-|\mathbf{r}| & (\mathbf{1}-\mathbf{r})\cdot \mathbf{n}^{(\nu)} & - \mathbf{r}\cdot(\mathbf{1}-\mathbf{n}^{(\nu)}) & - |\mathbf{r}|
\end{pmatrix}
\begin{pmatrix}
\delta_1 \\
\delta_2 \\
\delta_3 \\
\delta_4 
\end{pmatrix} = \mathbf{0} ,
\end{equation}
where $\nu$ is the number of elements of $S_\psi$. Regarding $\mathbf{\delta} = (\delta_1,\ldots,\delta_4)$ as a variable, we would like to know when there exists $\mathbf{r} \in \{0,1\}^N$ such that \eqref{eq.lineqsys} has a solution over $(\mathbb{R}^+)^4$, i.e.~a positive solution. Given such a solution, for any $l=1,\ldots,\nu$ we must have
\begin{equation}
\label{eq.rhsindep}
(\mathbf{1}-\mathbf{r})\cdot \mathbf{n}^{(l)} \, \delta_2 - \mathbf{r}\cdot(\mathbf{1}-\mathbf{n}^{(l)}) \delta_3 = |\mathbf{r}| \delta_4 - (N-|\mathbf{r}|) \delta_1 ,
\end{equation}
where the right-hand side is independent of $l$. Note that this condition can never be fulfilled if $\mathbf{r} = \mathbf{0}$ or $\mathbf{r} = \mathbf{1}$, because the two sides of the equation then have opposite signs. However, if the condition is satisfied, then for any pair $l,l'=1,\ldots,\nu$ we have
\begin{equation}
(\mathbf{1}-\mathbf{r})\cdot (\mathbf{n}^{(l)} - \mathbf{n}^{(l')}) \, \delta_2 - \mathbf{r}\cdot(\mathbf{n}^{(l')}-\mathbf{n}^{(l)}) \delta_3 = 0 .
\end{equation}
Hence, for a positive solution to exist, for each pair of support states either $\mathbf{n}^{(l)}$ and $\mathbf{n}^{(l')}$ have an equal number of 1's in positions where $\mathbf{r}$ has 0 and an equal number of 1's in positions where $\mathbf{r}$ has 1, or
\begin{equation}
\frac{\mathbf{r}\cdot(\mathbf{n}^{(l)}-\mathbf{n}^{(l')})}{(\mathbf{1}-\mathbf{r})\cdot (\mathbf{n}^{(l)} - \mathbf{n}^{(l')})} = - \frac{\delta_2}{\delta_3} < 0 ,
\end{equation}
is a negative constant independent of $l$, $l'$. On the other hand, if an $\mathbf{r} \neq \mathbf{0}, \mathbf{1}$ exists fulfilling these conditions, then a positive solution of \eqref{eq.lineqsys} is guaranteed to exist. This is because the left-hand side of \eqref{eq.rhsindep} is then independent of $l$ and thus one can always find positive $\delta_1$ and $\delta_4$ which make the equality true.

\section{Maximal filtering probability in GHZ-state machine}\label{GHZprob}

Naturally, since $N$ local filters are performed on the steady state of an $N$-qutrit autonomous thermal machine, the probability of a successful filtering decreases with $N$.  It is therefore reasonable to ask what this maximal possible success probability is. This can be determined analytically by considering the flow of population in the steady state of the GHZ machine.

Since a cold reset always takes a system out of the filtered sub-space, the maximal success probability is obtained in the limit $\gamma_h \gg \gamma_c$, i.e.~the opposite of the limit maximising the fidelity of the generated state with the target state. To determine $p_{\text{suc}}$ in this limit, let $S_k$ denote the set of all eigenstates of the joint free Hamiltonian where $k$ cold qutrits are in one of the excited states (all in the same one), while the remaining $N-k-1$ cold qutrits are in the ground state. For instance, in $S_{N-1}$ we have the states  $S_{N-1} = \{ \ket{0,\bar{1}}, \ket{1,\bar{1}}, \ket{2,\bar{1}}, \ket{0,\bar{2}}, \ket{1,\bar{2}}, \ket{2,\bar{2}}\}$ while $S_0$ consists of the states  $S_0 = \{ \ket{0,\bar{0}}, \ket{1,\bar{0}}, \ket{2,\bar{0}} \}$. We will compare the flows of population into and out of the $S_k$. However, first we argue that within each $S_k$, the populations on each of the states are equal in the steady state. We first note that all processes (the evolution driven by the $H_{\text{int}}$ of the GHZ machine, as well as hot and cold resets) are symmetric in the states $\ket{\bar{1}}$ and $\ket{\bar{2}}$ of the cold qutrits. The populations of states with the hot qutrit in a fixed state and a fixed number of cold qutrits excited to the same excited state, and which differ only in whether this state is $\ket{1}$ or $\ket{2}$ must therefore be equal in the steady state. In contrast, $H_{\text{int}}$ is not symmetric in the states $\ket{0}$, $\ket{1}$, $\ket{2}$ of the hot qutrit, and hence populations of states with the hot qutrit in different levels are not expected to be equal in the steady state in general. However, in the limit $\gamma_h \gg \gamma_c$, there are many hot resets between each cold one. This will then equalise the populations within each set $S_k$ before a cold reset causes a transition to $S_{k-1}$. Hence, all populations with each $S_k$ are equal in the steady state.

\begin{figure}[t]
	\begin{center}
		\includegraphics[width=0.6\linewidth]{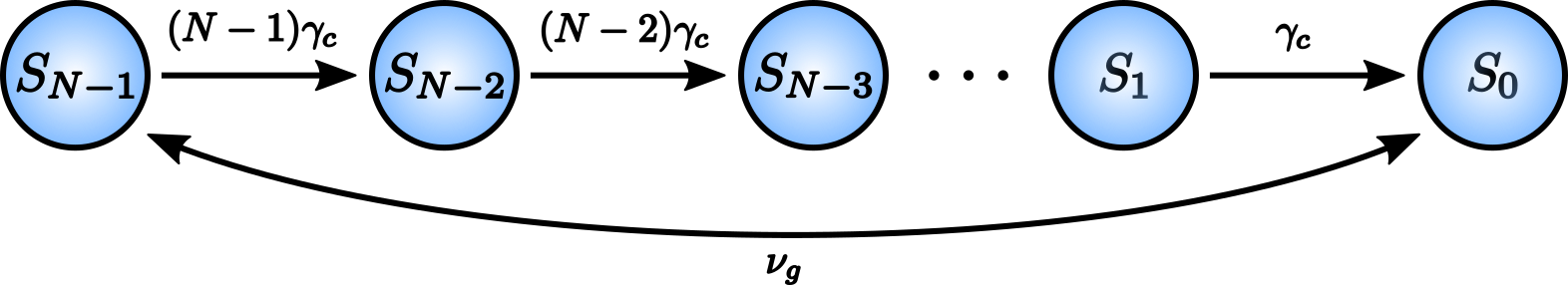}
		\caption{Flow diagram for population entering and leaving the sets of states $S_k$ with $k$ cold qubits excited. The rates per state in the set of origin are indicated.}
		\label{fig.psucflow}
	\end{center}
\end{figure}

We can now draw the flow diagram shown in \figref{fig.psucflow} for population transfer between the $S_k$. In the steady state, the flow into each set $S_k$ must equal the flow out. If we denote the population per state in $S_k$ by $P_k$, we therefore have, for $k=2,\ldots,N-1$
\begin{equation}
\label{eq.Poprelpsuc}
k \gamma_c |S_k| P_k = (k-1)\gamma_c |S_{k-1}| P_{k-1} .
\end{equation}
The number of states in the set $S_k$ is given by
\begin{equation}
\begin{split}
|S_0| & = 3 , \\
|S_k| & = 6 \begin{pmatrix} N-1 \\ k \end{pmatrix} , \hspace{0.75cm} k>0 .
\end{split}
\end{equation}
Inserting in \eqref{eq.Poprelpsuc} and rearranging, one finds that
\begin{equation}
\label{eq.poprel1}
\begin{split}
P_k & = \frac{k-1}{k} \begin{pmatrix} N-1 \\ k \end{pmatrix}^{-1} \begin{pmatrix} N-1 \\ k-1 \end{pmatrix} P_{k-1}  \\
& = \frac{k-1}{N-k} P_{k-1} , \hspace{0.75cm} k=2,\ldots,N-1 .
\end{split}
\end{equation}
From this it follows that $P_{N-1}=P_1$ and $P_{N-2} = P_2$ etc. That is,
\begin{equation}
P_{N-k} = P_{k} , \hspace{0.75cm} k=1,\ldots,N-1 .
\end{equation}
To determine the relation with $P_0$, we note that $H_{\text{int}}$ drives swaps between the states $\ket{2\bar{0}} \leftrightarrow \frac{1}{\sqrt{2}}(\ket{1\bar{1}} + \ket{0\bar{2}})$ and hence between $S_0$ and $S_{N-1}$. This process is a unitary rotation. Nevertheless, in the steady state it still results in a flow of population with a constant rate, which we can denote $\nu_g$. Focusing on the flow in and out of $S_{N-1}$, we can write
\begin{equation}
\nu_g P_{2\bar{0}} = \nu_g (P_{0\bar{2}} + P_{1\bar{1}}) + (N-1)\gamma_c |S_{N-1}| P_{N-1} .
\end{equation}
As argued above, when $\gamma_h \gg \gamma_c$, all states in each $S_k$ are equally probable, and so
\begin{equation}
\frac{1}{3} \nu_g P_0 = \frac{1}{3} \nu_g P_{N-1} + 6 (N-1)\gamma_c P_{N-1} .
\end{equation}
Now, if further $\nu_g \gg (N-1)\gamma_c$, then
\begin{equation}
\label{eq.poprel2}
P_0 = P_{N-1} .
\end{equation}
Finally, normalisation of the steady state requires that
\begin{equation}
\label{eq.poprel3}
1 = \sum_{k=0}^{N-1} |S_k| P_k = 3P_0 +  6 \sum_{k=1}^{N-1} \begin{pmatrix} N-1 \\ k \end{pmatrix} P_{k}
\end{equation}
Together, Eqns.~\eqref{eq.poprel1}, \eqref{eq.poprel2}, and \eqref{eq.poprel3} provide  $N$ independent equations from which the populations $P_k$, $k=0,\ldots,N-1$ can be determined. Explicitly, we can first express everything in terms of $P_0$. For $k \geq 1$
\begin{equation}
\label{eq.popspsuc}
P_k = \prod_{s=1}^{k-1} \frac{s}{N-s-1} P_1 = \begin{pmatrix} N-2 \\ k-1 \end{pmatrix}^{-1} P_0
\end{equation}
where we used that $P_1=P_{N-1}=P_0$. Then, from \eqref{eq.poprel3}
\begin{align}
1 & = \left[3 + 6 \sum_{k=1}^{N-1} \begin{pmatrix} N-1 \\ k \end{pmatrix} \begin{pmatrix} N-2 \\ k-1 \end{pmatrix}^{-1} \right] P_0 \\
& = \left[3 + 6 (N-1) \sum_{k=1}^{N-1} \frac{1}{k} \right] P_0 \\
& = 3\left[1 + 2 (N-1) h_{N-1} \right] P_0 ,
\end{align}
and hence
\begin{equation}
\label{eq.popP0psuc}
P_0 = \frac{1}{3\left[1 + 2(N-1)h_{N-1}\right]}
\end{equation}
where $h_n$ is the $n$'th harmonic number. We can now compute the probability for successful filtering, given the steady-state populations \eqref{eq.popspsuc} and \eqref{eq.popP0psuc}. The success probability becomes
\begin{align}\label{maxpsuc}
p_{\text{suc}} & = P(\text{hot qutrit not in $\ket{2}$},\text{no cold in $\ket{0}$}) \\
& = 4 P_{N-1} = 4 P_0 = \frac{4}{3\left[1 + 2(N-1)h_{N-1}\right]} \label{eq.psucanal} \\
& \approx \frac{4}{3 N \log(N)} ,
\end{align}
where the last line is valid for large $N$. We note that the assumption $\nu_g \gg (N-1)\gamma_c$ leading to \eqref{eq.poprel2} may not formally be justified for the local master equation. However, we have checked that the final expression \eqref{eq.psucanal} is consistent with solutions obtained for $N \leq 8$ without making this assumption.

It is interesting to observe that the critical $p_\text{suc}$ for obtaining a non-trivial GHZ-state fidelity approaches the above maximal value \eqref{maxpsuc} of $p_\text{suc}$ rapidly already for $N=3,4$ displayed in Fig.~\ref{GHZfidelityFig}. Provided that this observation extends to larger $N$, it is interesting to note that genuinely multipartite entanglement can be generated with a success probability which decreases only log-linearly with $N$.

\section{Nonlocality versus filtering probability in the GHZ-state and cluster-state machines}
\label{app.GHZClusternonlocality}

\begin{figure}
	\centering
	\begin{minipage}{0.49\textwidth}
	    \centering
    	\includegraphics[width=0.95\columnwidth]{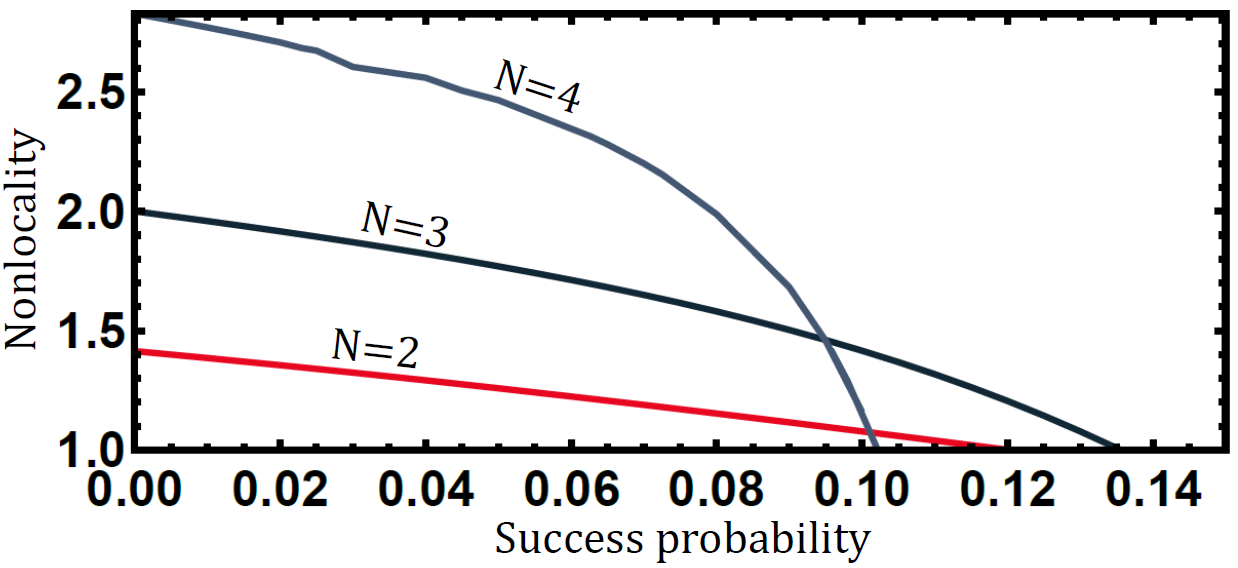}
	    \caption{Nonlocality versus filtering success probability for $N=2,3,4$ in a GHZ machine with one hot system and $N-1$ cold systems. The results are obtained numerically by optimising over $\gamma_h,g\leq 10^{-2}\Delta_\text{min}$.}\label{MerminFig}	
	\end{minipage}\hfill
	\begin{minipage}{0.49\textwidth}
		\centering
		\includegraphics[width=0.95\columnwidth]{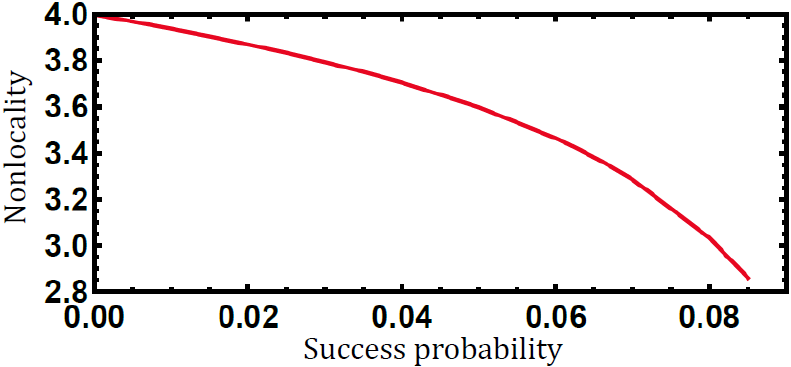}
		\caption{Nonlocality versus filtering success probability for the cluster state machine. The results are obtained by constrained optimisation over $\gamma_h,\gamma_c,g\leq 10^{-2}\Delta_\text{min}$.}\label{ClusterNonlocFig}	
	\end{minipage}
\end{figure}

A particularly strong form of entanglement is that which can violate a Bell inequality. Therefore, we have considered whether the states generated by the GHZ machine at fixed success probabilities have the ability of violating Bell inequalities. To this end, we have focused on the Mermin inequalities \cite{Mermin1990} which is a family of Bell inequalities applicable to scenarios in which $N$ observers share a state and perform one of two local measurements with binary outcomes. These inequalities are known to be maximally violated by a GHZ state. Let the input of the $k$'th observer in the Bell scenario be $x_k\in\{0,1\}$ and the corresponding output be $a_k\in\{0,1\}$.  We use a somewhat modified variant \cite{Tavakoli2016} of the Mermin inequalities which reads
\begin{eqnarray}
\frac{1}{2^N}\sum_{x_1\ldots x_N\in \{0,1\}} \left\lvert \left\langle\prod_{k=1}^{N} (A_0^{(k)}+(-1)^{x_k}A_1^{(k)}) \right\rangle \right\rvert \leq 1,
\end{eqnarray}
where 
\begin{equation}
\langle A_{x_1}^{(1)}\ldots A_{x_N}^{(N)}\rangle =\sum_{a_1\ldots a_N} (-1)^{a_1+\ldots +a_N}P(a_1\ldots a_N\lvert x_1\ldots x_N).
\end{equation}
We have fixed the measurements of each observer to be those required for a maximal violation with a GHZ state. For $N=2$, the optimal measurements are $\sigma_x$ and $\sigma_z$ for one observer, and $(\sigma_z+\sigma_x)/\sqrt{2}$ and $(\sigma_z-\sigma_x)/\sqrt{2}$ for the other observer. For $N=3$ we have let all three observers perform either $\sigma_x$ or $\sigma_y$, and for $N=4$ one observer performs either $\sigma_x$ or $\sigma_y$ whereas the remaining three choose between $(\sigma_x+\sigma_y)/\sqrt{2}$ and $(\sigma_x-\sigma_y)/\sqrt{2}$. We have numerically obtained the trade-off between nonlocality the filtering success probability. The results are illustrated in Fig.~\ref{MerminFig}. We conclude that the states generated by the GHZ machine can violate Bell inequalities for reasonable $p_{\text{suc}}$.

We have also performed an analogous analysis for the states generated at fixed success probabilities in the cluster-state machine. Specifically, we have considered whether these states can violate a Bell inequality tailored for cluster states  \cite{Scarani2005}. We have restricted ourselves to the measurements optimal for a cluster state $1/2(\ket{0000}+\ket{0011}+\ket{1100}-\ket{1111})$ which is unitarily equivalent to the target state \eqref{eq.cluster}. Hence, after a suitable local unitary, the Bell expression reads 
\begin{equation}
B=\langle \sigma_x\sigma_y\sigma_y\sigma_x+ \sigma_x\sigma_y\sigma_x\sigma_y+ \openone\sigma_z\sigma_x\sigma_x-\openone\sigma_z\sigma_y\sigma_y\rangle,
\end{equation}
which is bounded by $B\leq 2$ in all local hidden variable models. With a cluster state, one can achieve $B=4$. The trade-off between $B$ and the success probability of fitering is displayed in Fig.~\ref{ClusterNonlocFig}. We find that the generated states are nonlocal for any $p_\text{suc}$ up to its maximal value.

\section{Lindblad-type master equation}
\label{app.lindblad}

To demonstrate that our results are not restricted to the simple reset model employed in the main text, here we provide a Lindblad-type master equation, which can be derived from a microscopic model with bosonic baths. The reset model \eqref{mastereq}-\eqref{Lreset} is replaced by
\begin{equation}
\label{eq.lindblad}
\frac{d}{dt} \rho = -i [H_{\text{free}}+H_{\text{int}},\rho] + \sum_k \Gamma_k n_B(E_k,T_k) \mathcal{D}[A^+_k]\rho(t) + \sum_k \Gamma_k \left(1+n_B(E_k,T_k)\right) \mathcal{D}[A^-_k]\rho(t),
\end{equation}
where $\Gamma_k$ denotes the rate of a transition, $n_B(E,T) = 1/(e^{E/T}-1)$ is the Bose-Einstein distribution and $\mathcal{D}$ denotes the dissipator \footnote{We remark that this model suppresses one transition, in analogy with the Lindblad-type master equation considered in Ref.~\cite{Tavakoli2018}}.

\begin{figure}
	\centering
	\includegraphics[width=0.55\columnwidth]{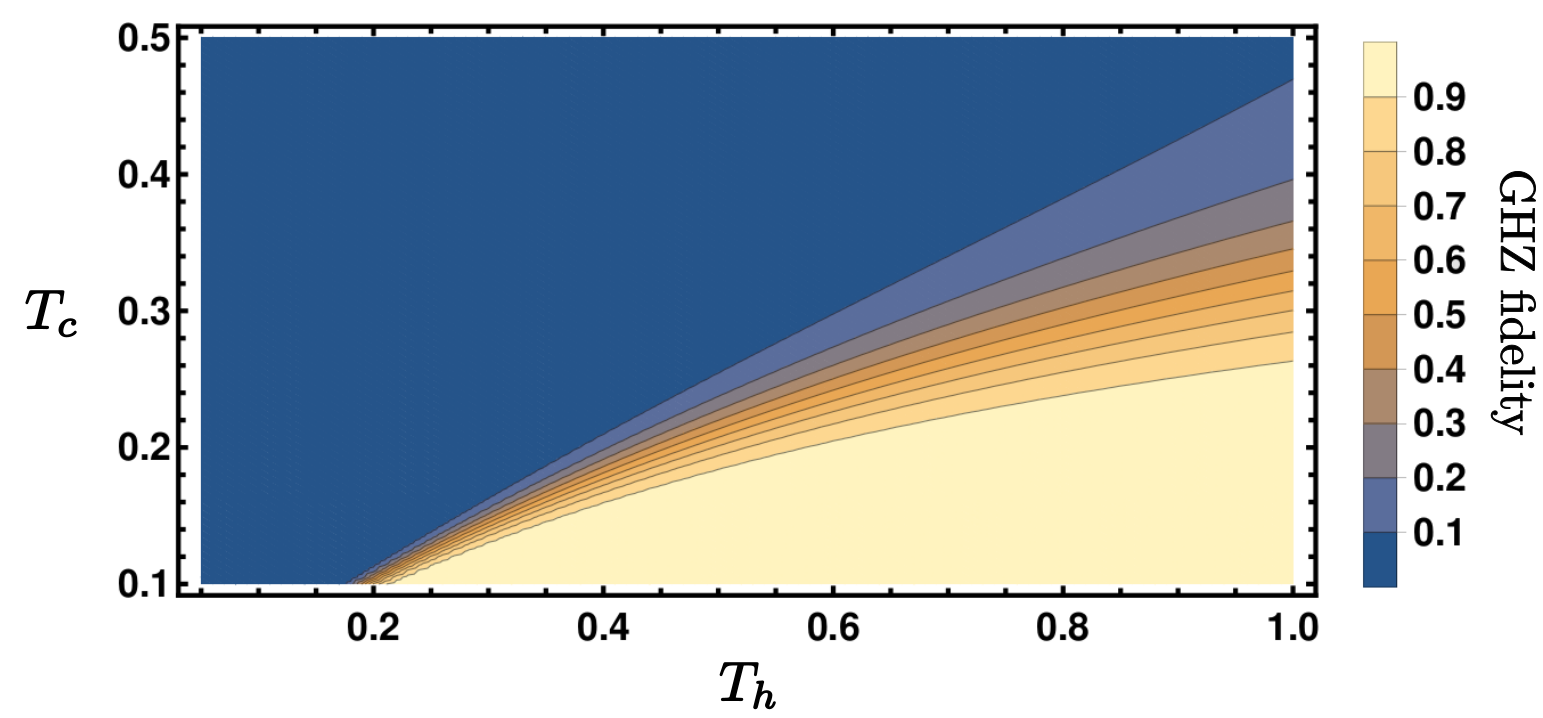}
		\caption{Fidelity of the filtered state with the GHZ state versus the bath temperatures when using the Lindblad-type master equation \eqref{eq.lindblad}. The plot is for $N=3$ parties and the parameter settings are $\Gamma_1=10^{-4}$, $\Gamma_2=\Gamma_3=5\times 10^{-3}$, $g=1.6\times 10^{-3}$, $\Delta^{(1)}=1$ $\Delta^{(2)}=2.5$.}\label{fig.GHZtempLindblad}
\end{figure}

Results from the Lindblad-type model qualitatively agree with those of the reset model. As an example, we again consider a GHZ target state for three parties ($N=3$), solve for the steady state, and find the GHZ fidelity of the filtered state as a function of $T_h$ and $T_c$. The result is shown in \figref{fig.GHZtempLindblad}. Just as in \figref{fig.GHZtemp} in the main text, we see that high fidelities can be attained with reasonably low temperature gradients. Parameter values are chosen based on recent experimental results in circuit-QED \cite{Pop2014,Jerger2016,Cottet2017,Lin2018}, see also \cite{Tavakoli2018}.

\end{document}